\documentstyle{article}
\date{\today}

\begin{document}
\title{The possible $\Sigma^0$-$\Lambda$ mixing in QCD Sum Rule}

\author{{Shi-lin Zhu,$^{1,2}$ W-Y. P. Hwang,$^3$ and Ze-sen Yang$^1$}\\
{$^1$Department of Physics, Peking University, Beijing, 100871, China}\\
{$^2$Institute of Theoretical Physics, Academia Sinica}\\
{P.O.Box 2735, Beijing 100080, China}\\
{$^3$Department of Physics, National Taiwan University, Taipei, 
Taiwan 10764}
}
\maketitle

\begin{center}
\begin{minipage}{120mm}
\vskip 0.6in
\begin{center}{\bf Abstract}\end{center}
{We calculate the on-shell $\Sigma^0$-$\Lambda$ mixing parameter $\theta$
with the method of QCD sum rule. 
Our result is $\theta (m^2_{\Sigma^0})
= (-)(0.5\pm 0.1) $ MeV. The electromagnetic interaction is not included.\\
{\large Keywords: $\Sigma^0$-$\Lambda$ mixing, QCD sum rules}\\
{\large PACS: 11.30.Hv, 14.20.Jn, 12.38.Lg  }\\
}
\end{minipage}
\end{center}

\large

\vskip 1.5cm
Isospin independence and charge symmetry is only approximate in the 
strong interaction. It is believed that the up and down quark mass 
difference and the electromagnetic interaction cause all the  
isospin violations \cite{hen1,rept}. Experimentally a strong signature 
for the $\rho^0$-$\omega$ mixing has been observed in the cross-section 
measurement of the reaction $e^+ e^-\to \pi^+ \pi^-$ \cite{aa}.
The effect from the electromagnetic interaction is of the opposite 
sign and much smaller than the experimentally determined 
$\langle \rho^0 |H_{\mbox{CSB}} |\omega \rangle $, while the current 
quark mass difference plays a dominant role \cite{bb}. 
Strong evidence for $\pi$-$\eta$-$\eta^{\prime}$ has also 
been obtained from studies of $\eta^{\prime}$ \cite{cc}, 
$\psi^{\prime}$ \cite{dd} and $\psi$ \cite{ee} decays. 
In this work we study the possible $\Sigma^0$-$\Lambda$ mixing with 
the method of QCD sum rules \cite{SVZ}. The $\rho^0$-$\omega$ mixing 
has been analysed within the same framework \cite{SVZ,hat}. The isospin 
symmetry breaking sources in such an approach come from the  
current quark mass difference $\delta m =m_u-m_d \ne 0$ and 
and the quark condensate difference $\gamma = {\langle 0| {\bar d}d |0 \rangle 
\over \langle 0| {\bar u}u | 0 \rangle } -1 \ne 0$. We do not 
take into account of the electromagnetic interaction in the present work. 
\par
We may study the $\Sigma^0$-$\Lambda$ mixing through 
the mixed propagator in the QCD vacuum
\begin{equation}
i \int d^4x e^{ipx} \langle 0|\Sigma^0 (x) {\bar \Lambda}(0) |0 \rangle 
=(-) {({\hat p} +m_{\Sigma^0})\theta ({\hat p}+m_{\Lambda})
\over (p^2 -m_{\Sigma^0}^2 +i\epsilon) 
(p^2 -m_{\Lambda}^2 +i\epsilon) } 
\end{equation}
where the mixing parameter $\theta (p^2)$ is defined through the following  
effective Lagragian:
\begin{equation}
L^{\mbox{mix}} = \theta ({\bar \Psi}_{\Sigma^0} \Psi_{\Lambda} +
{\bar \Psi}_{\Lambda} \Psi_{\Sigma^0})
\end{equation}
and may be measured through the decays 
like $\psi \to \Lambda {\bar \Sigma}^0
+ {\bar \Lambda}  \Sigma^0$ in the future experiment.
\par
In order to calculate the mixing parameter we study the two-point 
correlator at the quark level as:
\begin{equation}                                                       
\Pi (p)= i\int d^4 x \langle 0|T \{ \eta_{\Sigma^0}(x),{\overline  \eta}_{\Lambda}(0)\} 
|0\rangle e^{ip\cdot x}  \, ,
\end{equation}
The $\eta_{\Sigma^0}$ and $\eta_{\Lambda}$ 
are the currents with $\Sigma^0$ and $\Lambda$ quantum numbers
\begin{equation}
\eta_{\Sigma^0}(x) =\epsilon^{abc} {1\over \sqrt{2}} \{
[{u^a}^T (x) C\gamma_{\mu} d^b(x) ]\gamma_5 \gamma^{\mu}  s^c(x) 
+[{d^a}^T (x) C\gamma_{\mu} u^b(x) ]\gamma_5 \gamma^{\mu}  s^c(x) \}
\, ,
\end{equation}
\begin{equation}
\eta_{\Lambda}(x) =\epsilon^{abc} {\sqrt{2\over 3}} \{
[{u^a}^T (x) C\gamma_{\mu} s^b(x) ]\gamma_5 \gamma^{\mu}  d^c(x) 
-[{d^a}^T (x) C\gamma_{\mu} s^b(x) ]\gamma_5 \gamma^{\mu}  u^c(x) \}
\, ,
\end{equation}
where $u^a(x)$, $T$ and $C$ are the quark field, the transpose and the 
charge conjugate operators. $a$, $b$, $c$ is the color indices. 
The interpolating currents couple to the baryon states with the 
overlap amplititude $\lambda$. 
\begin{equation}
\langle 0| \eta_{\Sigma^0}(0) |\Sigma \rangle =\lambda_{\Sigma} \nu_{\Sigma} (p) \, ,
\end{equation}
\begin{equation}
\langle 0| \eta_{\Lambda}(0) |\Lambda \rangle =\lambda_{\Lambda} \nu_{\Lambda} (p) \, ,
\end{equation}
where the $\nu (p)$ is a Dirac spinor. 
\par
The correlation function $\Pi (p)$ may be expressed as  
\begin{equation}
\begin{array}{ll}
\label{c}
\langle 0| T\eta_{\Sigma^0}(x) {\bar \eta}_{\Lambda}(0) |0\rangle =
-{2\over \sqrt{3}} 
{\rm i} \epsilon^{abc}                  
\epsilon^{a^{\prime} b^{\prime} c^{\prime}}\{
& \gamma_5 \gamma^{\mu} S_s^{a a^{\prime}}(x)
\gamma_{\nu} C [S_u^{b b^{\prime}}(x) ]^T C \gamma_{\mu} 
S_d^{c c^{\prime}}(x) \gamma^{\nu}\gamma_5 \\
&
- \gamma_5 \gamma^{\mu} S_s^{a a^{\prime}}(x)
\gamma_{\nu} C [S_d^{b b^{\prime}}(x) ]^T C \gamma_{\mu} 
S_u^{c c^{\prime}}(x) \gamma^{\nu}\gamma_5 
\}
\end{array}
\end{equation}
where $iS^{ab}(x)$ is the quark propagator \cite{ykc}.  
\begin{equation}
\label{quark}
\begin{array}{ll}
{\rm i}S_q^{ab}(x) =&\langle 0|T[q^a(x) {\bar q}^b(0)]|0\rangle \\
 &=\frac{{\rm i}\delta^{ab}}{2\pi^2 x^4}{\hat x}
   +\frac{{\rm i}}{32\pi^2}\frac{\lambda^n_{ab}}{2}g_s G_{\mu\nu}^n \frac{1}{x^2}
   (\sigma^{\mu\nu}{\hat x}+{\hat x}\sigma^{\mu\nu})
   -\frac{\delta^{ab}}{12}\langle {\bar q} q\rangle         \\
 &+\frac{\delta^{ab} x^2}{192} \langle g_s {\bar q}\sigma \cdot G q\rangle
-\frac{ \langle {\bar q} q\rangle \langle g_s^2 G^2 \rangle x^4 }{2^9 \times 3^3} \delta^{ab} 
 -\frac{m_q \delta^{ab}}{4\pi^2 x^2} \\
 &+\frac{m_q}{32\pi^2}g_s\frac{\lambda_{ab}^n}{2} G^n_{\mu\nu} \sigma_{\mu\nu} \ln (-x^2)
 -\frac{\delta^{ab} \langle g^2_c G^2 \rangle }{2^9 \times 3\pi^2} m_q x^2 \ln (-x^2) \\
 &+\frac{i \delta^{ab} m_q \langle {\bar q} q\rangle }{48}{\hat x}
 -\frac{i m_q \langle g_s {\bar q}\sigma \cdot G q\rangle \delta^{ab} x^2 {\hat x} }{2^7\times 3^2}\\
 & +\cdots
\end{array}
\end{equation}
\par 
At the hadronic level 
\begin{equation}
\Pi (p) =(-) \lambda_{\Sigma^0} \lambda_{\Lambda}
{({\hat p}+m_{\Sigma^0})\theta (p^2) ({\hat p}+m_{\Lambda})
 \over (p^2 -m_{\Sigma^0}^2 +i\epsilon) 
(p^2 -m_{\Lambda^0}^2 +i\epsilon) } \, .
\end{equation}
\par
The diagrams with nonzero contribution are presented in Fig. 1. In the limit 
of exact isospin symmetry $\Pi (p)$ vanishes. There are two isospin 
symmetry breaking parameters $\delta m $ and $\gamma$. Each diagram is 
either propotional to $\delta m$ or to $\gamma$. 
We explicitly keep the current quark mass term in the quark propagator up 
to order $O(m_q)$, which is denoted by a cross. The current quark  
mass enters a diagram either through expanding the free quark propagator 
up to $O(m_q)$ or through the equation of motion. 
Since the strange quark mass is not small, we keep terms 
like $m_s(m_d -m_u)$.  
The calculation is standard as in the QCD sum rule analysis of the 
baryon mass. Equating the correlator $\Pi (p)$ 
at the quark level and 
$\Pi (p) $ at the hadronic level we arrive at two sum rules corresponding 
to two different structures $1$ and ${\hat p}$. 
Here we present the final result after 
Borel transformation. 
\par
For structure $1$:
\begin{equation}
\label{z1}
\begin{array}{ll}
{1\over \sqrt{3}}
& \{
 \delta m M^8_B E_3 L^{-\frac{8}{9}} 
 +\gamma a M_B^6 E_2 
 -{b \over 8} \delta m M_B^4 E_1 L^{-\frac{8}{9}} 
-{4\over 3}\delta m  a_s m_s M_B^4 E_1 L^{-\frac{8}{9}}  \\
&
- {1\over 3}\gamma m_s a a_s M_B^2 E_0 
+ {1\over 3}\delta m a^2 M_B^2 E_0 
-{1 \over 72} \gamma b a M_B^2 E_0 \\
&
+{7\over 48} \delta m a_s m_s m^2_0 M_B^2 E_0  L^{-\frac{38}{27}} 
-{5\over 8} \delta m a m^2_0  m_sM_B^2 \ln {M_B^2\over \mu^2}
E_0  L^{-\frac{38}{27}} 
\} \\
&= (2\pi )^4 \lambda_{\Sigma} \lambda_{\Lambda}
  e^{-\frac{m^2}{M_B^2}}
 2m^2 \theta (m^2_{\Sigma^0}) (1+A_1 M_B^2)
\end{array}
\end{equation}
\par
For structure ${\hat p}$:
\begin{equation}
\label{z2}
\begin{array}{ll}
-{1\over \sqrt{3}}
& \{
 \delta m  m_s M^6_B E_2 L^{-\frac{4}{3}} 
 +4 \gamma m_s a M_B^4 E_1 L^{-\frac{4}{9}} 
 +4 \delta m a_s M_B^4 E_1 L^{-\frac{4}{9}} 
+{2\over 3}\gamma  a a_s M_B^2 E_0 L^{\frac{4}{9}}  \\
&
-{1\over 4}\gamma  m_s a m^2_0 M_B^2 E_0  L^{-\frac{26}{27}} 
-{1\over 4} \delta m a_s m^2_0 M_B^2 E_0  L^{-\frac{26}{27}} 
+{1\over 8} \delta m a m^2_0 M_B^2 E_0  L^{-\frac{26}{27}} 
\} \\
&= (2\pi )^4 \lambda_{\Sigma} \lambda_{\Lambda}2 m
  e^{-\frac{m^2}{M_B^2}}
 \theta (m^2_{\Sigma^0}) (1+A_2 M_B^2)
\end{array}
\end{equation}
where $\delta m =m_d -m_u$ and $\gamma  = {\langle {\bar d }d \rangle \over 
\langle {\bar u }u \rangle  } -1$,  
$m={m_{\Sigma^0}+m_{\Lambda}\over 2}= 1.15\mbox{GeV}$ is the average mass. 
$m_s =150$MeV is the strange quark mass. 
$y=\frac{W^2}{M_B^2}$ and the factors,   
$E_n (y)=1-e^{-y} \sum\limits_{k=0}^{n} \frac{1}{k!} y^k $,  
are used to subtract the continuum contribution \cite{IOFFE}. 
$W^2=3.4$GeV$^2$ is the continuum threshold which is 
determined together with the overlap amplititude 
$(2\pi )^4\lambda_{\Sigma}^2 =1.88$GeV$^6$,  
$(2\pi )^4\lambda_{\Lambda}^2 =1.64$GeV$^6$  
from the mass sum rules \cite{REINDERS,HW}. 
We adopt the ``standard'' values for the various  
condensates 
$b=\langle 0| g_s^2 G^2 |0 \rangle =0.474$GeV$^4$, 
$a=-(2\pi )^2 \langle 0 | {\overline  u}  u |0 \rangle 
=0.55   \mbox{GeV}^3 $, 
$a_s=-(2\pi )^2 \langle 0 | {\overline  s}  s |0 \rangle 
=0.55  \times 0.8 \mbox{GeV}^3 $, 
$a m_0^2 =(2\pi )^2 g_s \langle 0 | {\overline  u}\sigma\cdot G  u |0 \rangle $, 
$m_0^2 =0.8\mbox{GeV}^2$.
$L=\frac{\ln (\frac{M_B}{\Lambda_{\mbox{QCD}}})}{\ln (\frac{\mu}{\Lambda_{\mbox{QCD}}})}$, 
$\Lambda_{\mbox{QCD}}$ is the QCD parameter, $\Lambda_{\mbox{QCD}}= 100$MeV, 
$\mu = 0.5$GeV is the normalization point to which the used values of 
condensates are referred. 
\par
We further improve the numerical analysis by taking into account of the 
renormalization group evolution of the sum rules (\ref{z1}) and (\ref{z2}) 
through the anomalous dimensions of the various condensates and currents. 
$A_1$ and $A_2$ are constants to be determined from the sum rule.  
They arise from 
the mixing with the excited states ${\Sigma^0}^{\ast}$-$\Lambda$ or 
$\Sigma^0$-$\Lambda^{\ast}$ which were first 
introduced in the QCD sum rules analysis of the nucleon magnetic 
moments \cite{IOFFE}.
The working interval for the Borel mass 
$M_B^2$ is $1.3 \mbox{GeV}^2 \leq M_B^2 \leq 2.5 \mbox{GeV}^2$ where both the 
continuum contribution and power corrections are controllable.
Moving the factor $(2 \pi )^4 \lambda_{\Sigma} \lambda_{\Lambda}
e^{- {m^2\over M_B^2}}$
on the right hand side to the left and 
fitting the new sum rule with a straight line approximation we may extract 
the mixing parameter $\theta$. 
\par
Various theoretical approaches 
\cite{rept,ykc,gl,wein,bar,nar,dom}
yield consistent results for the quark mass difference, 
$\delta m =3.2 \pm 0.4$ MeV. 
And the difference of the up and down quark condensate 
has been analyzed with the chiral perturbation theory 
\cite{gl}, the QCD sum rules for scalar and pseudoscalar mesons 
\cite{nar,dom,nar2}, effective models of QCD incorporating the dynamical 
breaking of chiral symmetry \cite{paver,hat2}, and the QCD sum rules for 
baryons \cite{ykc}. The numerical results from the above approaches are  
$\gamma = -(6-10) \times 10^{-3}$ \cite{gl}, 
$-(10 \pm 3) \times 10^{-3}$ \cite{nar}, 
$-(7-9) \times 10^{-3}$ \cite{paver,hat2} 
and $-6.57 \times 10^{-3}$ \cite{ykc}.
\par 
In both of the sum rules (\ref{z1}) and (\ref{z2})
the contribution from $\delta m$ and $\gamma$ 
has opposite sign. 
The mixing parameter from (\ref{z1}) and (\ref{z2}) would contradict each other 
if $\gamma$ is too large or too small. In Fig. 2 the Borel mass dependence    
of the mixing parameter and the fitting straight line is shown. Through the 
intersection with the Y-axis, we can obtain the value of $\theta$ directly.
With $\gamma = -(7 \pm 1) \times 10^{-3}$ 
and $\delta m =3.0\pm 0.4$ MeV,  
our final result is $\theta (m_{\Sigma^0}^2) 
= (-)(0.5\pm 0.1)$MeV. It can be seen from Fig. 2 that (\ref{z1}) and 
(\ref{z2}) yields almost the same value for $\theta$.  
In summary we have calculated  
the on-shell $\Sigma^0$-$\Lambda$ mixing parameter, which may be measured 
in the future experiment.

\par
This work is supported in part by the Postdoctoral Science Foundation 
of China and 
the National Natural Science Foundation
of China. It is also supported in part by the National Science Council of 
R.O.C. (Taiwan) under the grant NSC84-2112-M002-021Y.

\newpage
\begin{center}
{\bf Figure Captions}
\end{center}
\par
\noindent
Fig. 1 Relevant diagrams in the QCD sum rules analysis of the 
$\Sigma^0$-$\Lambda$ mixing up to dimension seven.  
The current quark mass correction is denoted by a cross.  
If two crosses appear in the same diagram, one of them 
comes from the strange quark.
\vskip 1cm
\par
\noindent
Fig. 2 (a) The Borel mass dependence of the mixing parameter. The 
solid  
curve is the QCD sum rule prediction for 
$\gamma =-7 \times 10^{-3}$ 
from equation (\ref{z1}) 
for the structure $1$ with $\delta m= 3.0$MeV
after the numerical factor $(2 \pi )^4 \lambda_{\Sigma^0}^2\lambda_{\Lambda}^2
e^{- {{m^2\over M_B^2}}}$ is moved to the left hand
side. The dotted curve is the fitting straight line. 
The Borel mass $M_B^2$ is in unit of GeV$^2$. 
The mixing parameter $\theta$ is unit of MeV.\\ 
(b) The Borel mass dependence of the mixing matrix for the structure 
${\hat p}$. Notations are the same as in (a). 

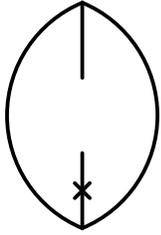
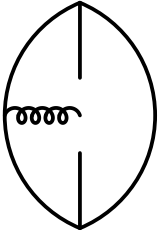
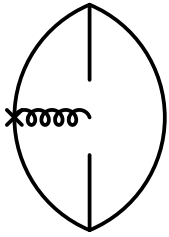
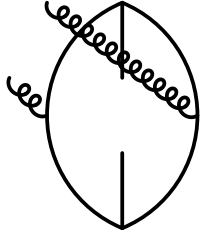
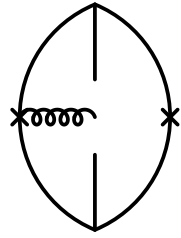
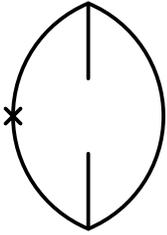
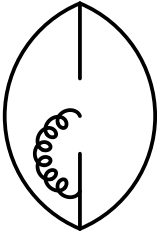
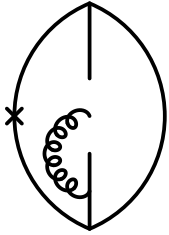
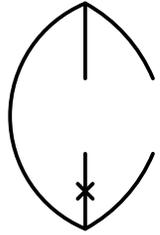
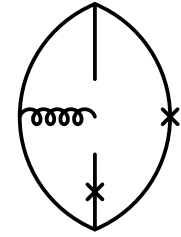
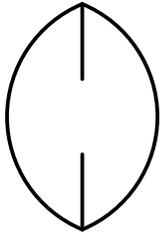
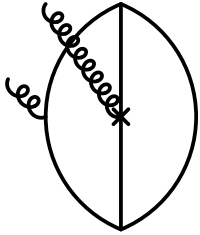
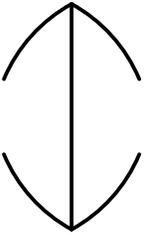
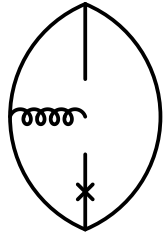
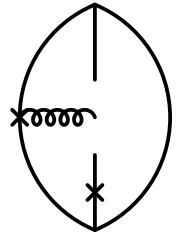
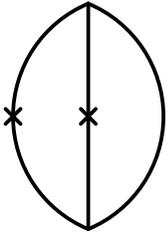
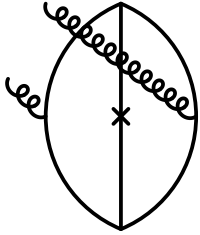
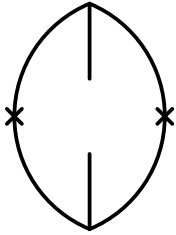
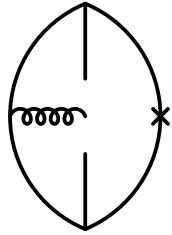
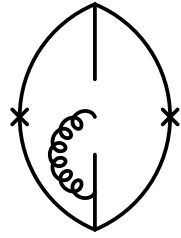
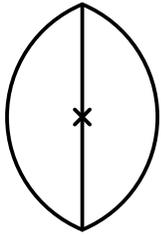
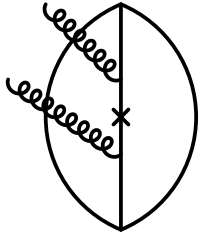
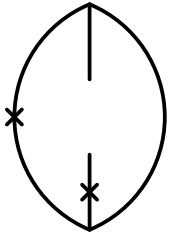
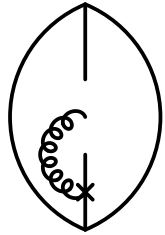
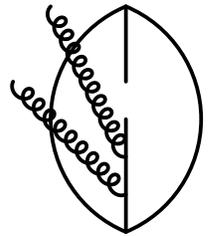
\end{document}